\documentstyle[12pt]{article}
\newcommand{\ncm}{\newcommand}
\ncm{\oH}{\bar{H}}
\ncm{\us}{\quad\mbox{using}\quad}
\ncm{\ra}{\rightarrow}
\ncm{\ot}{\otimes}
\ncm{\DH}{D(\oH)}
\ncm{\TH}{T(\oH)}
\ncm{\bea}{\begin{eqnarray}}
\ncm{\eea}{\end{eqnarray}}
\ncm{\beas}{\begin{eqnarray*}}
\ncm{\eeas}{\end{eqnarray*}}

\ncm{\im}{\imath}
\ncm{\ba}{\begin{array}}
\ncm{\ea}{\end{array}}
\ncm{\ul}{\underline}
\ncm{\ol}{\overline}

\def\hoek{\hbox{\vrule height 2.5ex depth 0pt \vrule width 2.5ex height .4pt
 depth 0pt}}
\def\haak#1#2{
\mathop{\hoek\llap{\vbox to 2.5ex{ \vfil
\hbox{$\scriptstyle#1$\hskip 2.8ex} \vfil}}}
\limits_{#2} }
\def\hook#1#2{\setbox0=\hbox{$\scriptstyle#1$}
\hskip\wd0\haak{\box0}{#2}}
\ncm{\str}{\rule{0cm}{3.5mm}}
\ncm{\om}{\omega}
\ncm{\ep}{\epsilon}
\ncm{\be}{\begin{equation}}
\ncm{\ee}{\end{equation}}
\ncm{\al}{\alpha}
\ncm{\bt}{\beta}
\ncm{\gm}{\gamma}
\ncm{\dl}{\delta}
\ncm{\varep}{\varepsilon}
\ncm{\zt}{\zeta}
\ncm{\et}{\eta}
\ncm{\th}{\theta}
\ncm{\kp}{\kappa}
\ncm{\lm}{\lambda}
\ncm{\rh}{\rho}
\ncm{\hl}{\hline}
\ncm{\sg}{\sigma}
\ncm{\ta}{\tau}
\ncm{\ph}{\phi}
\ncm{\phv}{\varphi}
\ncm{\ch}{\chi}
\ncm{\ps}{\psi}
\ncm{\nn}{\nonumber}
\title{Quantum symmetries in discrete gauge theories}
\author{F.Alexander Bais\thanks{email: bais@phys.uva.nl},
$\,$ Peter van Driel\thanks{email: vandriel@phys.uva.nl}
$\,$ and
Mark de Wild Propitius\thanks{email: mdwp@phys.uva.nl}
\\ Instituut voor Theoretische Fysica\\Valckenierstraat 65\\
1018XE Amsterdam}
\date{December 1991}
\begin{document}
\maketitle
\begin{abstract}
We analyse the fusion, braiding and scattering properties of discrete
non-abelian anyons. These occur in (2+1)-dimensional theories where a
gauge group $G$ is
spontaneously broken down to some discrete subgroup $H$. We identify the
quantumnumbers of the electrically and magnetically charged sectors of the
remaining discrete gauge theory, and show that on the quantum level the
symmetry group $\bar{H}$ is extended to the (quasi-triangular) Hopf algebra
$D(\bar{H})$. A conjugacy class paired with a centralizer representation forms
an irreducible representation. The fusion rules for arbitrary discrete
non-abelian anyons are completely determined by the nontrivial comultiplication
(i.e. tensorproduct) of this algebra. It allows for a clear interpretation of
Cheshire charge. Also the braid-matrix ${\cal R}$, which determines
the statistical properties and the two particle Aharonov-Bohm scattering,
is fixed and satisfies the Yang-Baxter equation. Most of our considerations
are relevant for discrete gauge theories in (3+1)-dimensional space time as
well.
\end{abstract}
\vspace*{3cm}
Preprint ITFA-91-40 to be published in  Physics Letters B.
\newpage  \noindent
{\bf Introduction.}
In two spatial dimensions particles or excitations with (fractional) charge and
(fractional) flux may have exceptional spin-statistics properties. These
so-called {\em anyons} are of particular interest because of their
possible relevance
in the description of the fractional quantum Hall effect, and their
appearance in models for two dimensional superconductivity. So far,
most attention has been given to the  {\em abelian} anyons
\cite{wil1}. In this paper we focus on the non-abelian variety, which
appear as excitations in gauge theories,
where some continuous non-abelian group $G$
is spontaneously broken down to a finite non-abelian group $H$ by a
nonvanishing expectation value of some Higgs field~$\Phi$.
In $(2+1)$ dimensions this leads to particle-like topological
excitations which carry a non-abelian magnetic flux, labelled by
elements of the homotopy group $\pi_1(G/H)\simeq\pi_1(\bar{G}/\bar{H})\simeq
\bar{H}$, where $\bar{G}$ is the covering group of $G$ and $\bar{H}$
the lift of $H$ in $\bar{G}$. There has been a recent increase of
interest in these peculiar objects and many interesting though
partial results on their fundamental properties have been
obtained~\cite{wil2,pre1,wil3}.
In this paper we propose a general quantum description of these magnetic-
and all the other (anyonic) excitations in this discrete
$\bar{H}$ gauge theory. To be explicit, we give a
treatment of their fusion,
braiding and scattering properties, based on an extension of the
residual group $H$ to a new symmetry algebra $D(\bar{H})$.
The results reported here have been obtained by
exploiting a striking analogy with certain aspects of string- and
conformal field theory (notably orbifold models), and consequently, also
with topological field theory.\\
For convenience, we have restricted ourselves to (2+1)-dimensional space time,
but it should be stressed that most of our considerations are relevant for
discrete gauge theories in (3+1) dimensions as well. In the latter case the
magnetic excitations, labelled by $\pi_1(G/H)$, are string-like.\\[.5cm]
{\bf The model.}
For the sake of concreteness we will restrict our considerations to the
case where $G=SO(3)$, but our analysis applies in general. Consider the
Lagrangian
\be
{\cal L}= -\frac{1}{4}F_{\mu\nu}^a F^{\mu\nu}_a + (D_\mu \Phi)^*\cdot(D^\mu
\Phi) - V(\Phi),    \label{action}
\ee
where $A_\mu^a$ $(\mu = 0,1,2; a= 1,2,3)$ is the Yang-Mills potential
and $F_{\mu\nu}^a$ the corresponding fieldstrength. We take the covariant
derivative as $D_\mu \Phi = (\partial_\mu +ieA_\mu^a T_a)\Phi$ , with
$T_a$ the (hermitean) generators of $SO(3)$ in the representation
of the Higgs field $\Phi$.
\\[.5cm]
{\bf Topological flux sectors.}
With a suitable choice for the representation of $\Phi$
and its invariant potentential $V(\Phi)$ one may break $SO(3)$ down
to any discrete subgroup $H$, i.e. a cyclic- $Z_N$, a dihedral- $D_N$ or
a pointgroup like $T, O$ or $Y$. The broken phase supports
topological excitations with magnetic fluxes  labelled by the
elements $h$ of the double group $\bar{H}$. Although regular time
independent finite energy solutions can be constructed, we are here only
interested in their long range behaviour, which is fixed by minimising the
three terms in the energy density seperatedly. From $|F|^2 = 0$ we
conclude that $F=0$, so
$A$ becomes locally a pure gauge. Minimising $V(\Phi)$ tells us that $\Phi$
should take  values in the vacuum manifold $\Phi= \Phi_0 \in G/H$.
Finally, the condition $D\Phi=0$ can be integrated along a closed path
$\gamma$ in space to yield the condition
\be
{\cal P}\exp \int_\gamma A_i^a(s) T_a ds^i = \Gamma(h) \;\;\;\;\,\,\, h\in H,
\ee
where ${\cal P}$ denotes path ordering and $\Gamma$ an $SO(3)$ representation.
This condition states that the holonomy is restricted to lie in
$H$~\cite{bai1}. Indeed, after taking the limit where the symmetry breaking
scale (and hence all masses) go to infinity, one is basically left with (dually
charged) pointlike excitations which interact purely topologically through
locally flat gauge connections with holonomy restricted to the discrete group
$H$. All the statements we will make pertain to the theory in the
aforementioned limit, thus effectively  describing the long range interactions
between topologically charged particles.\\
Under the residual gauge group $\bar{H}$ the magnetic charges $h$ transform by
conjugation
\be \label{gaugeaction}
h \longmapsto bhb^{-1},
\ee
with $b \in \bar{H}$. This
suggests that we should think of the conjugacy classes $C$ as degenerate
multiplets. This will indeed turn out to be the case.
\\[.5cm]
{\bf Fusion and braiding of fluxes.}
Let us consider the {\em fusion} and {\em braiding}
properties of the magnetic excitations.
The classical fusion- or composition rule for two fluxes $h$ and $k$
corresponds to the group multiplication $h\cdot k$, which indicates
the importance of the ordering if $\bar{H}$ is non-abelian.
Indeed, if we simply braid $h$ and $k$ by interchanging them such that
$k$ passes behind $h$ then $k$ suffers from a {\em metamorphosis}:
\be \label{metamorphosis}
|h> |k> \longmapsto |hkh^{-1}\!>|h>,
\ee
so the ordered product remains invariant~\cite{bai1,tou1}. This nontrivial
braiding property has important consequences on the quantum level. Clearly, if
we want to discuss the quantum statistical properties of particles carrying
these fluxes, we have to diagonalise the braid-matrix ${\cal R}$ on
the two-particle hilbertspace, thus we have to consider the vectorspace spanned
by the group elements, the so-called {\em group algebra} ${\cal C}(\bar{H})$.
One expects that already at this level,
${\cal R}$ may have eigenvalues different from
$\pm 1$, meaning that such particles have anyonic properties and exhibit
Aharonov-Bohm scattering~\cite{wil2}.\\
The group $\bar{H}$ acts by conjugation on ${\cal C}(\bar{H})$,
which is  reducible under this action.
In fact, even the conjugacy classes generally do not
form irreducible representations. It is important to note that there are now
two notions of fusion:
\begin{itemize}
\item  the tensor product of representations in ${\cal C}(\bar{H})$
(i.e. the ordinary representation ring of $\bar{H}$),
\item  the class algebra obtained by multiplication of all
the elements in the two classes.
\end{itemize}
Surprisingly enough, these two notions are not compatible (see for
instance the discussion after equation (\ref{fus})),
which suggests that either the classical notion of flux
composition, or the implementation of the $\bar{H}$ symmetry needs
adjustment. This problem is nicely resolved in the generalized
framework that we are to propose.
\\[.5cm]
{\bf Electric charges.} First of all, we know that in the magnetic vacuum
sector, the
electric charges will correspond to representations of the residual
symmetry group $H$. If we add fields in representations of the
original gauge group $SO(3)$ (or $SU(2)$), then we know how they decompose into
a
direct sum of irreducible $H$ (or $\bar{H}$) representations.
However, in a sector with nontrivial flux one can only implement
global transformations
belonging to the centralizer $^A\!N\!\subset\!H$ (or $\bar{H}$)
of the group
element $h$, which labels the magnetic charge~\cite{bal,pre1,wil3}. The
centralizer is obviously a subgroup of $H$ (or $\bar{H}$), and depends only
on the class $^A\!C\!\subset\!\bar{H}$ to which $h$ belongs.
So in general, we should think of
flux/charge sectors labelled as follows:
\be
|\,^A\!C, \,^{\alpha}\Gamma>,   \label{state}
\ee
where $\,^{\alpha}\Gamma$ labels the $\alpha$-th irreducible representation
of the centralizer $\,^A\!N$. These sectors are spanned by the states
$\{|^A\!g_j,\,^{\alpha}\!v_i>\}_{j=1,\ldots,k}^{i=1,\ldots,
\mbox{\scriptsize dim}\alpha}$, with the basis elements of
the representations $^{\alpha}\!\Gamma$ of
$^A\!N$ denoted by $^{\alpha}\!v_i$. Clearly, the number of sectors is finite.
\footnote{Note that the sum of the squares of the dimensions of the
representations equals the {\em squared} order of $\bar{H}$:
$\sum_{A,\alpha} |\,^A\!C|^2|\,^{\alpha}\!\Gamma|^2=\sum_{A} |\,
^A\!C|^2|\,^A\!N|=|\oH|^2.$ As we will see later, this is the order
of the extended symmetry algebra.}\\
We can assign a spin $s$ to the dyonic sectors
of the theory~\cite{tho1} (by, for example calculating the conserved
angular momentum of the classical solution).
The spin factor $w(A,\alpha)$ for (\ref{state}) is related to the
character $\alpha(\,^A\!g_1)$ in the representation $^{\alpha}\!\Gamma$:
\be
w(A,\alpha)=\exp 2\pi i s = \alpha(\,^A\!g_1)/\alpha(e),    \label{spinfactor}
\ee
with $\alpha(\,^A\!g_1)=\mbox{Tr}\,^{\alpha}\!\Gamma(^A\!g_1)$.
In models in which the conventional
spin-statistics connection is realized (such as
the abelian case), we have the identity $w(A,\alpha)=\exp i\theta (A,\alpha)$,
where $\theta$ denotes the statistical parameter typically equal to
the product of the charge $q$  and the flux $\phi$.
As we will see later, the relation between spin and statistics is more
involved in the non-abelian case.\\[.5cm]
{\bf The lattice of charge/flux sectors.}
To illustrate the situation and facilitate forthcoming
discussions, we start with the diagram of
admissable electric charges and magnetic fluxes in a compact
$U(1)$ theory (Figure 1). Both of them are quantised (say $q=ne$ and
$\phi=2\pi m/e$, obeying the Dirac condition), and the lattice
extends indefinitely in both directions. If we consider
this $U(1)$ as a subgroup of $SU(2)$, then the charges can be
half integral ($q=ne/2$) and consequently the flux should
double  ($\phi=4\pi m/e$). However, in two spatial dimensions
there are no topological flux sectors in the unbroken
$SU(2)$, because its fundamental group is trivial (i.e.
$m=m$ mod 1). In the group $SO(3)\simeq SU(2)/Z_2$, one has again
that $q=ne$ and $\phi= 2\pi m/e$ where now $m=m$ mod $2$ so
that only one nontrivial magnetic sector remains. If we now
break $SO(3)$ down to some discrete group $H$,
we can determine the magnitude of the flux $h$ by looking at $\bar{H}$
or better at its centralizer $^h\!N \subset\bar{H}$. In most cases
this is some group $Z_{2M}$.
It is clear that the charges will be
$q=ne/2$ with $n=n$ mod $2M$ and $\phi=4\pi m/2Me$ with $m=m$ mod $2M$ for
$SU(2)$, or alternatively $q=ne$ with $n=n$ mod $M$ and $\phi=2\pi m/Me$ with
$m=m$ mod $2M$ in $SO(3)$.
We see that the charge/flux lattice becomes
periodic, but the periodicity in the electric direction depends on the
magnitude of the flux.
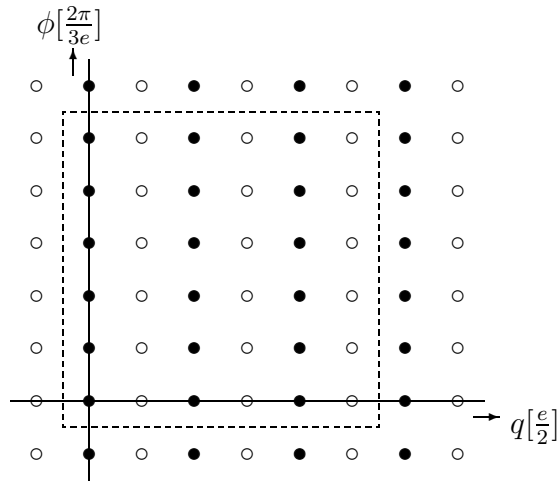
\begin{figure}
\setlength{\unitlength}{0.7mm}
\begin{center}
\begin{picture}(100,80)(-15,-15)
\put(-5,-5){\dashbox(60,60)[t]{}}
\put(-15,0){\line(1,0){90}}
\put(0,-15){\line(0,1){80}}
\thinlines
\multiput(-10,-10)(0,10){8}{\multiput(0,0)(10,0){9}{\circle{1.8}}}
\multiput(0,-10)(0,10){8}{\multiput(0,0)(20,0){4}{\circle*{2.2}}}
\put(-3,62){\vector(0,1){5}}
\put(73,-3){\vector(1,0){5}}
\put(-10,70){$\phi[\frac{2\pi}{3e}]$}
\put(80,-6){$q[\frac{e}{2}]$}
\end{picture}
\end{center}
\caption{The charge/flux lattice for $\bar{H}=Z_6$. In $SO(3)$ only the
charges denoted by the filled circles occur.}
\end{figure}
One should note that for a non-abelian group $H$,
the reduction of the allowed charges/fluxes to a two-dimensional lattice
is of course an oversimplification, if for example more then one particle
is in question
(as in the case of braiding and fusion). This is one of the
reasons to develop some more algebraic apparatus in the
following sections. \\[5mm]
{\bf The symmetry algebra $D(\bar{H})$.}
Once we have identified the superselection sectors~(\ref{state})
in our discrete gauge theory, we should answer the following
questions: How do we implement the $\bar{H}$ symmetry? And consequently, what
are the fusion braiding and scattering properties of the spectrum?
Specifically, we have to define the action of an arbitrary
group element $\bar{h}\in \bar{H}$ in any centralizer
representation $\,^{\alpha}\!\Gamma$, and we need to develop a
tensorcalculus for excitations~(\ref{state}), which involves
somehow `tensoring' representations of different centralizers. A similar
problem is encountered with holomophic orbifolds in conformal field
theory~\cite{dvvv}.
As was shown by Dijkgraaf, Pasquier and Roche, the solution amounts to
an extension of the symmetry group $\bar{H}$ to the quasi-triangular
Hopf algebra $D(\bar{H})$. We summarize some of
its characteristics~\cite{dpr1}. For a more general discussion of Hopf algebras
in
connection with conformal field theory, we refer to~\cite{dri1}.\\
$\DH$ is a finitely generated algebra of order $|\oH|^2$. A basis is
given by $\{ \hook{g}{x} \}_ { g,x \in \oH}$ ,
with multiplication
\be   \label{product}
\hook{g}{x}\cdot\hook{h}{y}=\delta_{g,xhx^{-1}}\hook{g}{xy}.
\ee
Here, $\delta_{g,h}=1$ if $g=h$ and $0$ otherwise.
As a sidenote, we mention that the subgroup $\oH$ is generated by the elements
\be  \label{subgroup}
\hook{1}{x}\equiv\sum_{g}\hook{g}{x}.
\ee
Note that for fixed $g$, and $x,y\in \,^g\!N$, the product
(\ref{product}) reduces to the product in the subgroup $^g\!N$.
The full representation theory of $\DH$ is obtained by
inducing the representations of these subgroups~\cite{dpr1}.
Let $\{\,^A\!C\}$ be the set of conjugacy classes of $\oH$ and
introduce a fixed but arbitrary ordering $^A\!C=\{^A\!g_1,
\;^A\!g_2,\ldots,\,^A\!g_k\}$. Let $^A\!N$ be the centralizer of
$^A\!g_1$ and $\{^A\!x_1,\,^A\!x_2,\ldots,\,^A\!x_k\}$ be a set of
representatives of the equivalence classes of $G/^A\!N$, such that
$^A\!g_i=\,^A\!x_i\,^A\!g_1\,^A\!x_i^{-1}$ and choose $^A\!x_1=e$. Consider
the complex vectorspace $V^A_{\alpha}$ spanned by the  basis
$\{|^A\!g_j,\,^{\alpha}\!v_i\!>\}_{j=1,\ldots,k}^{i=1,\ldots, \mbox{\scriptsize
dim}\alpha}$.
We denote the basis elements of the unitary
irreducible representation $^{\alpha}\!\Gamma$
of $^A\!N$ by
$^{\alpha}\!v_i$. This vectorspace carries a representation $\Pi^A_{\alpha}$ of
$D(\oH)$, given by
\be \label{13}
\Pi^A_{\alpha}(\,\hook{g}{x})|\,^A\!g_i,\,^{\alpha}\!v_j \!>=
\delta_{g,x\,^A\!g_ix^{-1}}|x\,^A\!g_ix^{-1},
\,^{\alpha}\!\Gamma(\,^A\!x_k^{-1}x\,^A\!x_i)\,^{\alpha}\!v_j \!>,
\ee
with $\,^A\!x_k$  defined by $\,^A\!g_k=x\,^A\!g_ix^{-1}$.\\
The set $\{\Pi^{A}_{\alpha}\}$ provides the complete set of
representations~\cite{lus1}.
Note that this set coincides with the spectrum~(\ref{state}) of the $\oH$
discrete gauge theory, as discussed in the previous paragraph. What's more;
$\DH$  gives rise to consistent (and partly anticipated) fusion rules
and braid properties of the particles in this spectrum.
\begin{description}
\item{\bf Fusion.} The fusion rules contain information about all possible
invariant couplings between the different particles in the spectrum, they are
as such relevant in the computation of non-elastic
scattering cross-sections. In order to
incorporate the action of $\DH$ on these couplings, one formally needs to
extend the action of $\DH$ to $\DH\otimes\DH$. This is done by means of the
comultiplication $\Delta$. For the algebra at hand,
the comultiplication reads as
\be  \label{comultiplication}
\Delta(\,\hook{g}{x})=\sum_{\stackrel{h,k\in
G}{hk=g}}\hook{h}{x}\ot\hook{k}{x}.
\ee
Since $\Delta$ is an algebra-morphism, the tensor product of two
representations $\Pi^A_{\al}$ and $\Pi^B_{\beta}$ is again  a
representation. This representation is in general  not
irreducible, but rather gives rise to the decomposition
\be\label{fusionrule}
\Pi^A_{\al}\otimes\Pi^B_{\beta}=N^{AB\gamma}_{\alpha\beta C}
\Pi^C_{\gamma},
\ee
where $N^{AB\gamma}_{\alpha\beta C}$ is the multiplicity of the irreducible
representations $\Pi^C_{\gamma}$. Relation (\ref{fusionrule}) is called a
fusion rule of the algebra $\DH$. The fusion algebra is a commutative
associative algebra and can therefore be diagonalized by a single matrix, $S$.
For $\DH$, this matrix takes the following form
\be
S^{AB}_{\alpha\beta}=\sum_{\stackrel{\,^A\!g_i\in\,^A\!C\,,^B\!g_j\in\,
^B\!C}{[\,^A\!g_i,\,^B\!g_j]=e}}
\alpha^*(\,^A\!x_i^{-1}\,^B\!g_j\,^A\!x_i)\beta^*(\,^B\!x_j^{-1}\,^A\!g_i\,
^B\!x_j).
\ee
This so-called `modular' $S$ matrix plays a profound role in conformal field
theory. It contains all the information about the fusion algebra \cite{ver0}:
\be
N^{AB\gamma}_{\alpha\beta C}=\sum_{D,\delta}\frac{
S^{AD}_{\alpha\delta}S^{BD}_{\beta\delta}
(S^{*})^{CD}_{\gamma\delta}}{S^{eD}_{0\delta}}.
\ee
It is related to charge conjugation through $S^2=C$.
In the present context, the $S$ matrix realises a higher-order
electric-magnetic duality on the space of states.

\item{\bf Braiding.}
The universal $R$-matrix on $\DH$ reads
\be
R=\sum_{g\in G} \hook{g}{e} \otimes \hook{1}{g}.\label{Rmatrix}
\ee  To obtain its action, we consider the two-particle state
$|\,^A\!g_i,\,^{\alpha}\!v_j\!> |\,^B\!g_k,\,^{\beta}\!v_l\!>\in
\!|\,^A\!C,\,^{\alpha}\!\Gamma>\!\otimes |\,^B\!C,\,^{\beta}\!\Gamma>$.
Using~(\ref{Rmatrix}), we define
\be\label{braidmatrix}
{\cal R}^{\alpha\beta}_{AB}\equiv
\sigma\circ(\Pi_{\alpha}^A\otimes\Pi_{\beta}^B)(R),
\ee
where $\sigma$ is the permutation operator.
The operator ${\cal R}$ implements a (positively oriented)
interchange of two particles.
Explicitly, on the state
$|\,^A\!g_i,\,^{\alpha}\!v_j\!>|\,^B\!g_k,\,^{\beta}\!v_l\!>$
the braid operation reads
\be \label{braidaction}
{\cal R}^{\alpha\beta}_{AB}|^A\!g_i,\,^{\alpha}\!v_j\!>\!
|^B\!g_k,\,^{\beta}\!v_l\!>=|\,^A\!g_i\,^B\!g_k\,^A\!g_i^{-1},
\,^{\beta}\!\Gamma(^B\!x_m^{-1\;A}\!g_i^B\!x_k)^{\beta}\!v_j\!> |^A\!g_i,
^{\alpha}\!v_j\!>,
\ee
where $^B\!x_m$ is defined through $^B\!g_m=\,^A\!g_i\,^B\!g_k\,^A\!g_i^{-1}$.
A few remarks are in order at this stage:
\begin{itemize}
\item the braiding of two pure magnetic fluxes gives the flux
metamorphosis~(\ref{metamorphosis}),
\item encircling a pure electric charge around a pure magnetic flux $^A\!g_j$
(a process effectuated by ${\cal R}^{\alpha 0}_{eA}{\cal R}^{0\alpha}_{Ae}$),
boils down to a transformation of the electric charge with
$^{\alpha}\!\Gamma(^A\!g_j)$. A result that also concides with earlier
observations~\cite{wil3},
\item  consistency of the braid operation ${\cal R}$ with the
extended symmetry $\DH$ is assured since ${\cal R}$ commutes with the
$\DH$ that is defined on the tensor product through the
comultiplication (\ref{comultiplication}),
\item  ${\cal R}$ satisfies the Yang-Baxter equation ${\cal R}_1{\cal R}_2
{\cal R}_1={\cal R}_2{\cal R}_1{\cal R}_2$. Here ${\cal R}_1$ acts on the
three-particle states as ${\cal R}\otimes {\bf 1}$ and ${\cal R}_2$ as ${\bf
1}\otimes {\cal R}$,
\item the braid-matrix is related to the modular $S$ matrix:
$ S=\frac{1}{|\bar{H}|}\mbox{Tr}({\cal R}^2)$.
\end{itemize}
\item{\bf Braid statistics.}
The ${\cal R}$-matrix commutes with the comultiplication
$\Delta$, so states which have the same eigenvalues under ${\cal R}$
also form representations of $D(\bar{H})$, and vice versa. Since
in the fusion product of a representation with itself different
representations can occur, the braid matrix generically has
distinct eigenvalues.
According to our earlier observation (\ref{spinfactor}) we can associate to
each representation
of $D(\bar{H})$ a spinfactor $w(A,\alpha)$. The phases
$m^{AB\gamma}_{\alpha\beta C}$ that appear in the monodromy matrix
${\cal R}^2$,
depend for two representations ($\Pi^{A}_{\alpha}$ and $\Pi^{B}_{\beta}$) on
the channel ($\Pi^{C}_{\gamma}$) allowed for by the fusionrule of these
representations. One  obtains the following formula \cite{dpr1}:
\be
m^{AB\gamma}_{\alpha\beta C} = \frac{w(C,\gamma)}{w(A,\alpha)w(B,
\beta)}\;. \label{braidfactor}
\ee
The conventional spin-statistics connection \cite{fro1} is retrieved from
$m^{A\bar{A}0}_{\alpha\bar{\alpha} e}$, where $(\bar{A},\bar{\alpha})$ denotes
the charge conjugated sector of $(A,\alpha)$.
\item{\bf Aharonov-Bohm scattering.}
The cross sections of elastic two-particle Aharonov-Bohm scattering
are completely determined by the monodromy matrix ${\cal R}^2$. The
explicit relation can be cast in the following form~\cite{ver1}
\be \label{Aharonov}
\frac{d\sigma}{d\varphi}=\frac{1}{2\pi k
\sin^2(\varphi/2)}\;\frac{1}{2}\,[1-\mbox{Re}<\psi_{in}|{\cal
R}^2|\psi_{in}>].
\ee Here $|\psi_{in}>$ denotes the incoming two-particle state, $k$
the relative wave vector. Besides the conventional abelian
Aharonov-Bohm scattering \cite{ahab} and the non-abelian pure flux
scattering \cite{wil2}, it can also be verified that this formula
governs scattering of pure electric charge $\Gamma$ off pure
magnetic flux $^A\!C$. In the latter case, the $\oH$ multiplet $\Gamma$
indeed decomposes into
multiplets of the centralizer $^A\!N$ of $^A\!C$, such that different
particles in the same centralizer multiplet acquire the same
Aharonov-Bohm phase~\cite{wil3}. \end{description}
{\bf An example: the quaternion group $\bar{H}=\bar{D}_2$}. Consider the
example of the double dihedral group $\bar{D}_2$. This is a group of order 8,
with five conjugacy classes, denoted $e,
\bar{e},X_1,
X_2$ and $X_3$. Table 1a exhibits the elements of these
conjugacy classes together with their centralizers. Table 1b
and 1c are the charactertables of the occurring centralizers.
\begin{center}
\begin{tabular}[t]{||l|c||} \hline
$ \mbox{Conjugacy class} \str                         $&$\str \mbox{Centr.}
   $\\ \hl
$ e={\bf 1} \str                                      $&$ \bar{D}_2$\\ \hl
$ \bar{e}=-{\bf 1}\str                                $&$ \bar{D}_2$\\ \hl
$ X_1=\{\imath\sigma_1,-\imath\sigma_1\}\str          $&$ Z_4 $\\ \hl
$ X_2=\{\imath\sigma_2,-\imath\sigma_2\}              $&$ \str Z_4 $\\ \hl
$ X_3=\{\imath\sigma_3,-\imath\sigma_3\}              $&$ \str Z_4 $\\ \hl
\end{tabular}
\begin{tabular}[t]{||l|r|r|r|r|r||} \hline
$\str \bar{D}_2$&$ e $&$ \bar{e}$&$ X_1  $&$ X_2  $&$ X_3  $\\ \hl
$ ^0\!\Gamma\str$&$ 1 $&$ 1      $&$ 1  $&$ 1  $&$ 1  $\\ \hl
$ ^1\!\Gamma\str$&$ 1 $&$ 1      $&$ 1  $&$-1  $&$-1  $\\ \hl
$ ^2\!\Gamma\str$&$ 1 $&$ 1      $&$-1  $&$ 1  $&$-1  $\\ \hl
$ ^3\!\Gamma\str$&$ 1 $&$ 1      $&$-1  $&$-1  $&$ 1  $\\ \hl
$ ^4\!\Gamma\str$&$ 2 $&$ -2     $&$ 0  $&$ 0  $&$ 0  $\\ \hl
\end{tabular}
\begin{tabular}[t]{||l|r|r|r|r|r||} \hline
$\str Z_4$&$ e $&$ g       $&$ g^2$&$ g^3   $\\ \hl
$ ^0\!\hat{\Gamma}\str$&$ 1 $&$ 1       $&$ 1  $&$ 1     $\\ \hl
$ ^1\!\hat{\Gamma}\str$&$ 1 $&$ \imath  $&$-1  $&$-\imath$\\ \hl
$ ^2\!\hat{\Gamma}\str$&$ 1 $&$ -1      $&$ 1  $&$-1     $\\ \hl
$ ^3\!\hat{\Gamma}\str$&$ 1 $&$ -\imath $&$-1  $&$\imath $\\ \hl
\end{tabular}\\
\begin{center}
{\em Table 1a.\qquad\qquad\qquad\qquad\qquad Table 1b.
\qquad\qquad\qquad\qquad Table 1c.}
\end{center}
\end{center}
The representations of $D(\bar{D}_2)$ are labelled as follows
\begin{center}
\vspace{-10mm}
\begin{minipage}[t]{3.8cm}
\beas
1        &\simeq& |e,\,^0\! \Gamma>,      \\
J_a      &\simeq& |e,\,^a\! \Gamma>,      \\
\phi     &\simeq& |e,\,^4\! \Gamma>,
\eeas
\end{minipage}
\begin{minipage}[t]{3.8cm}
\beas
\bar{1}        &\simeq& |\bar{e},\,^0\!\Gamma>,      \\
\bar{J}_a      &\simeq& |\bar{e},\,^a\!\Gamma>,      \\
\bar{\phi}     &\simeq& |\bar{e},\,^4\!\Gamma>,
\eeas
\end{minipage}
\begin{minipage}[t]{3.8cm}
\beas
\sigma_a^+&\simeq& |X_a,\,^0\!\hat{\Gamma}>,\\
\sigma_a^-&\simeq& |X_a,\,^2\!\hat{\Gamma}>,
\eeas
\end{minipage}
\begin{minipage}[t]{3.8cm}
\beas
\tau_a^+&\simeq& |X_a,\,^1\!\hat{\Gamma}>,\\
\tau_a^-&\simeq& |X_a,\,^3\!\hat{\Gamma}>,
\eeas \end{minipage}
\end{center}
where $a$ runs from $1$ to $3$. These representations constitute the complete
set of inequivalent irreducible representations of $D(\bar{D}_2)$. In
particular, we have 22 representations, divided among 8 1-dimensional- and 14
2-dimensional representations. Indeed, this adds up to the dimension of the
algebra: $8.1^2+14.2^2=8^2$.\\ The sets $\{J_a,\phi\}$ and $\{\bar{1},
\sigma^+_a\}$ respectively, form the pure electric- and pure magnetic
excitations in our discrete $\bar{D}_2$ gauge theory. The other
representations are dyonic excitations.\\ Using the algebraic machinery of
$D(\bar{D}_2)$ we can compute the $S$-matrix (Table 2) and
subsequently the fusion rules. We discuss some of these.
\begin{center}
\begin{tabular}{||l|l|r|r|r|r|r|r|r|r|r||} \hline
$\rule{0cm}{3.5mm}$&$1 $&$\bar{1} $&$J_a $&$\bar{J}_a $&$\phi $&$\bar{\phi}
$&$\sigma^+_a$ &$\sigma^-_a $&$\tau^+_a $&$\tau^-_a $\\ \hl
$\rule{0cm}{3.5mm}1$&$1 $&$1 $&$1 $&$1 $&$2 $&$2 $&$2 $&$2 $&$2 $&$2 $\\ \hl
$\rule{0cm}{3.5mm}\bar{1}$&$1 $&$1 $&$1 $&$1 $&$-2 $&$-2 $&$2 $&$2 $&$-2 $&$-2
$\\ \hl
$\rule{0cm}{3.5mm}J_b$&$1 $&$1 $&$1 $&$1 $&$2 $&$2 $&$2\ep_{ab} $&$2\ep_{ab}
$&$2\ep_{ab}
$&$2\ep_{ab} $\\ \hl
$\rule{0cm}{3.5mm}\bar{J}_b$&$1 $&$1 $&$1 $&$1 $&$-2 $&$-2 $&$2\ep_{ab}
$&$2\ep_{ab} $&$-2\ep_{ab}
$&$-2\ep_{ab} $\\ \hl
$\rule{0cm}{3.5mm}\phi$&$2$&$-2$&$2$&$-2$&$4 $&$-4 $&$0 $&$0 $&$0 $&$0 $\\ \hl
$\rule{0cm}{3.5mm}\bar{\phi}$&$2$&$-2$&$2$&$-2$&$-4$&$4  $&$0 $&$0 $&$0 $&$0
$\\ \hl
$\rule{0cm}{3.5mm}\sigma^+_b$&$2 $&$2 $&$2\ep_{ab} $&$2\ep_{ab} $&$0 $&$0
$&$4\delta_{ab} $&$-4\delta_{ab} $&$0 $&$0 $\\ \hl
$\rule{0cm}{3.5mm}\sigma^-_b$&$2 $&$2 $&$2\ep_{ab} $&$2\ep_{ab} $&$0 $&$0
$&$-4\delta_{ab}     $&$4\delta_{ab} $&$0 $&$0 $\\ \hl
$\rule{0cm}{3.5mm}\tau^+_b$&$2 $&$-2$&$2\ep_{ab} $&$-2\ep_{ab} $&$0 $&$0 $&$0
$&$0 $&$-4\delta_{ab} $&$4\delta_{ab} $\\ \hl
$\rule{0cm}{3.5mm}\tau^-_b$&$2 $&$-2$&$2\ep_{ab} $&$-2\ep_{ab} $&$0 $&$0 $&$0
$&$0 $&$4\delta_{ab} $&$-4\delta_{ab} $\\ \hl
\end{tabular}\\[.4 mm]
{\em Table 2. The modular S-matrix of $\bar{D}_2$, $\epsilon_{ab}= 1$ if $a=b$
and $-1$ otherwise.}
\end{center}
The fusion of the pure electric charges is dictated by the $\bar{D}_2$
representation ring \[
J_a\times J_a=1,\qquad
J_a\times J_b=J_c,\qquad
\phi \times \phi= 1 + \sum J_a,
\]
while amalgamation of electric charges with magnetic fluxes may yield dyonic
excitations
\[
J_a\times \sigma^+_a=\sigma^+_a,\qquad
J_a\times \sigma^+_b=\sigma^-_b,\qquad
\phi \times \sigma^+_a= \tau^+_a+\tau^-_a.
\]
Remarkably, the fusion of two pure magnetic fluxes seems to give rise to
electric charge creation, as for example in the fusion product of
$\sigma^+_a$ with itself
\be  \label{fus}
\sigma^+_a\times \sigma^+_a=1+J_a+\bar{1}+\bar{J}_a.
\ee
Before fusion this electric charge was present in the form of
so-called non-localizable Cheshire charge~\cite{pre1} (i.e. the
non-trivial
representation of the gauge group $\bar{D}_2$) carried by the magnetic
flux pair. In terms of these Cheshire charges we have
\be   \label{fus1}
(\,^0\!\Gamma+\,^1\!\Gamma)\times(\,^0\!\Gamma+\,^1\!\Gamma)
=
\,^0\!\Gamma+\,^1\!\Gamma+\,^0\!\Gamma+\,^1\!\Gamma.
\ee
Of course, charge creation is not the proper expression in this context.
A similar argument holds for the
common  centralizer $Z_4=\{e,X_a,\bar{e}, \bar{X}_a \}$.
In terms of the representations of this group, the
fusion (\ref{fus}) reads
\[
(\,^0\!\hat{\Gamma}+\,^0\!\hat{\Gamma})\times(\,^0\!\hat{\Gamma}+\,
^0\!\hat{\Gamma})
=
\,^0\!\hat{\Gamma}+\,^0\!\hat{\Gamma}+\,^0\!\hat{\Gamma}+\,^0\!\hat{\Gamma}.
\]
So, there is no electric charge creation at any level of symmetry
$D(\bar{D}_2) \supset \bar{D}_2 \supset Z_4$.\\
We also note that the class algebra is still
respected as an overall selection rule.
In (\ref{fus}), the associated class multiplication yields $X_a * X_a  = 2
e + 2 \bar{e}$. Clearly, the interpretation of the magnetic fluxes in terms of
$\bar{H}$ representations due to (\ref{gaugeaction}) yields a result that is
incompatible with (\ref{fus1}). (See the discussion in the paragraph `Fusion
and braiding of fluxes'.)\\
Finally, there is also the peculiar occurrence of the vacuum in the fusion of
for instance $\tau^{+}_a$ with itself
\be \label{fus2}
\tau^+_a \times \tau^+_a=1+J_a+\sum_{b\neq a} \bar{J}_b.
\ee
This is different from the usual situation where charge conjugation relates
complex conjugated (electric) representations. This naively relates $\tau^+_a$
to
$\tau^-_a$. However, due to the interplay between magnetic and electric
quantum numbers, this relation is altered. Indeed, in this example we find
$S^2=1$. \\
It is also instructive to determine the Aharonov-Bohm
cross-sections~(\ref{Aharonov}) for the elastic scattering processes in this
model.
We shall do this for a simple, yet nontrivial, example:
scattering  a $\bar{\phi}$ particle
off a $\tau_1^+$ particle.  Assume this two particle system to be in the
following four component quantumstate
\[
|\psi_{in}>=|\bar{e},\cos\theta\,^4\!v_1+\sin\theta\,^4\!v_2>
            |\cos\theta' \,X_1 + \sin\theta'\,\bar{X}_1,\,^{1}\!v>.
\]
Under  encircling the $\tau_1^+$ particle with the $\bar{\phi}$
particle this quantumstate transforms as
\beas
{\cal R}^2 |\psi_{in}>&=&
                {\cal R}^2 (|\bar{e},\cos\theta\,^4\!v_1+\sin\theta\,^4\!v_2>
                            |\cos\theta' X_1,\,^1\!v>\\
     & &\,\,\,\,\,
+|\bar{e},\cos\theta\,^4\!v_1+\sin\theta\,^4\!v_2>
                             |\sin\theta'\bar{X}_1,\,^1\!v>)\\
     &=&    |\bar{e},\cos\theta\,^4\!\Gamma(X_1)\,^4\!v_1
                      + \sin\theta\,^4\!\Gamma(X_1)\,^4\!v_2>
            |\cos\theta' X_1,\,^1\!\hat{\Gamma}(\bar{e})\,^1\!v>\\
     & &\,\,\,\,\,\,  +|\bar{e},\cos\theta \,^4\!\Gamma(\bar{X}_1)\,^4\!v_1
                 +\sin\theta\,^4\!\Gamma(\bar{X}_1)\,^4\!v_2>
            |\sin\theta'\bar{X}_1,\,^1\!\hat{\Gamma}(\bar{e})\,^1\!v> \\
     &=&    |\bar{e},\imath\cos\theta\, ^4\!v_2+\imath\sin\theta\,^4\!v_1>
            |\cos\theta' X_1 - \sin\theta'\bar{X}_1,-\,^1\!v>.
\eeas
In the last equality we used Table 1, in particular the fact that the
representation $^4\!\Gamma$ involves the pauli matrices:
$^4\!\Gamma(X_1)=-\,^4\!\Gamma(\bar{X}_1) = \imath\sigma_1$ .
The real part of the inproduct $<\psi_{in}|{\cal R}^2 |\psi_{in}>$ vanishes.
Plugging this result in~(\ref{Aharonov}) yields the following cross-section
\be
\frac{d\sigma}{d\varphi}(\bar{\phi},\tau^+_1)=\frac{1}{2\pi k
\sin^2(\varphi/2)}\,\,\frac{1}{2}.
\ee
The calculations of the other cross-sections, although in some cases
more involved, are completely analogous.\\[.5 cm]
{\bf Concluding remarks.}
In this paper we proposed a general framework to describe quantum properties
of discrete $\bar{H}$ gauge theories, based on the Hopf algebra $\DH$. It
incorporates among others an unified description of the Aharonov-Bohm
scatterings, fusion, and a clear interpretation of
Cheshire charges in this model. In a forthcoming paper~\cite{bai2},
we discuss the inclusion of a Chern-Simons term in
the action. We show that this corresponds to the introduction of a
non-trivial 3-cocycle on the algebra $\DH$. The representation theory, and
consequently the physical properties like fusion, braiding and scattering are
modified. In so doing we will argue that the C-S parameter is not just
quantized,
but also periodic. Furthermore, a correspondence with the
discrete topological field theories introduced by Dijkgraaf and Witten
emerges.\\[.5cm]
{\bf Acknowledgements.}
This work has been partially supported by the Dutch Science Organisation
FOM/NWO. We thank R. Dijkgraaf and E. and H. Verlinde, for clarifying
discussions on their work. We are also indebted to Aad Pruisken for his
support.

\end{document}